# THE LOCALIZED QUANTUM VACUUM FIELD

D. Dragoman – Univ. Bucharest, Physics Dept., P.O. Box MG-11, 077125 Bucharest, Romania, e-mail: danieladragoman@yahoo.com

ABSTRACT

A model for the localized quantum vacuum is proposed in which the zero-point energy of the quantum electromagnetic field originates in energy- and momentum-conserving transitions of material systems from their ground state to an unstable state with negative energy. These transitions are accompanied by emissions and re-absorptions of real photons, which generate a localized quantum vacuum in the neighborhood of material systems. The model could help resolve the cosmological paradox associated to the zero-point energy of electromagnetic fields, while reclaiming quantum effects associated with quantum vacuum such as the Casimir effect and the Lamb shift; it also offers a new insight into the Zitterbewegung of material particles.

INTRODUCTION

The zero-point energy (ZPE) of the quantum electromagnetic field is at the same time an indispensable concept of quantum field theory and a controversial issue (see [1] for an excellent review of the subject). The need of the ZPE has been recognized from the beginning of quantum theory of radiation, since only the inclusion of this term assures no first-order temperature-independent correction to the average energy of an oscillator in thermal equilibrium with blackbody radiation in the classical limit of high temperatures. A more rigorous introduction of the ZPE stems from the treatment of the electromagnetic radiation as an ensemble of harmonic quantum oscillators. Then, the total energy of the quantum electromagnetic field is given by $E = \Sigma_{k,s} \hbar w_k (n_{ks} + 1/2)$, where $n_{ks}$ is the number of quantum oscillators (photons) in the ($k$,$s$) mode that propagate with wavevector $k$ and frequency $w_k = |k|c = kc$, and are characterized by the polarization index $s$. The ZPE of the quantum field, $E_0 = \Sigma_{k,s} \hbar w_k / 2$, corresponds to a quantum state with no photons, which is called for this reason quantum vacuum. Although the vacuum carries the definite, non-fluctuating ZPE, the expectation values of the electric and magnetic fields in the vacuum state vanish (they fluctuate with zero mean). Quantum field theory assumes that the entire universe is bathed in ZPE.

This introduction of the ZPE is problematic in itself because the electromagnetic field with a certain frequency, seen as a many-photon quantum state, is mathematically equivalent to a harmonic oscillator with a formal unit mass. On the other hand, relativity theory states that the photon is massless; in particular, this implies that the photon cannot be stopped, as does a harmonic oscillator with finite mass at the position of its highest potential energy. Although only the many-photon quantum state and not a single photon is mathematically similar to a quantum harmonic oscillator, in that the photon number $n_{k,s}$ is analogous to the quantum number that labels the different excited states of a harmonic oscillator with mass, this ambiguous conceptual issue constitutes the first of many paradoxes associated to quantum vacuum. Note for completeness that the wavefunction of a single photon is not the same as the wavefunction of the first excited state of a quantum oscillator (see [2,3] and the references therein); it can be meaningfully defined if a relaxed definition of localizability of quantum systems is employed.

Another disturbing feature of the ZPE is that the total energy diverges in even finite volumes. More precisely, the vacuum free space energy density, equal to $(\hbar/2\pi^2 c^3)\int \omega^3 d\omega$ (the index $k$ of the frequency $\omega_k$ is removed here and in the remaining of the paper for notational simplicity), is infinite unless a cut-off frequency is arbitrarily introduced as an upper limit to the integral. Although renormalization procedures can take care of these mathematical infinities, the situation is quite unsatisfactory. Moreover, the contribution of the ZPE at the energy density in general relativity leads theoretically to a contribution to the effective cosmological constant (from the corresponding energy-momentum tensor) that is with at least 40 orders of magnitude higher than that estimated from experimental observations. The term "cosmological paradox" was coined to describe this situation; no theory can presently explain it in a satisfactory manner [4]. This contradiction between theory and experiment suggests that the quantum theory that predicts the existence of the ZPE in the outer space must be seriously re-examined.

Despite these considerations, the ZPE is a useful concept in elucidating several phenomena, which include the stability of quantum systems (in the sense that ZPE has an essential role in preserving the commutation relations of quantum systems interacting with the vacuum), the spontaneous emission of radiation, the Lamb shift, and the Casimir effect. Due to the recent advancements in nanotechnologies, the Casimir effect in particular has received a great deal of interest and has been subject to several experimental tests (see [5] and the review in [6], which includes both recent experimental and theoretical advances in the Casimir effect). It is caused by the boundary dependence of the ZPE and predicts the mutual attraction or repulsion of micro- and nano-sized objects depending on their geometry, size, dielectric constants, the topology and quality of the boundary. All experiments demonstrate that the Casimir force exists and that its dependence on the distance between the attracting objects is the same as predicted by the quantum theory. Therefore, the existence of the ZPE as described by standard quantum theory is experimentally established.

The aim of the present paper is to offer an explanation of the origin of the ZPE that is consistent with the experiments revealing its existence (in particular, experiments revealing the existence of the Casimir effect) and that at the same time could help elucidating the cosmological paradox. The resolution to the mystery of the ZPE is in its localization: the quantum vacuum does not exist throughout the universe, but only in the neighborhood of material systems. We show that the quantization of electromagnetic radiation in the form of an ensemble of harmonic oscillators is not required to explain the existence of the ZPE.

THE LOCALIZED QUANTUM VACUUM

Our model of quantum vacuum is based on the observation that the ZPE is indispensable for the stability of material systems. Therefore, we consider a particle with mass $m$ that can be modeled as a quantum harmonic oscillator with resonant frequency $w_0$. The eigenvalues of the Hamiltonian operator

$$\hat{H} = \hat{p}^2/2m + mw_0^2\hat{r}^2/2 \tag{1}$$

expressed in terms of momentum and coordinate operators $\hat{p}$ and $\hat{r}$, respectively, are $E_n = \hbar w_0(n+1/2)$, with $n = 0, 1, 2,\ldots$, and the quantum eigenstates are denoted by $|\Psi_n\rangle$. If the particle is in an excited state, say the first excited state with $n = 1$, it can spontaneously jump to the ground state by emitting a photon with energy $\hbar w$. The transition takes place with energy and momentum conservation, the emitted photon and the recoiled atom in the ground state being entangled. Atom recoil during spontaneous transitions to the ground state has been experimentally observed [7], and the wavefunction of the photon–ground state atom has been analytically computed in [3] in terms of a single-photon state $|1_k\rangle$ with a wavefunction in the coordinate representation

$$\Psi_{ph}(\boldsymbol{r}_{ph},t)\rangle = L^{-3/2} \sum_{\boldsymbol{k},s} C_{\boldsymbol{k},s} \boldsymbol{e}_{\boldsymbol{k},s} \exp[(i\boldsymbol{k}\cdot\boldsymbol{r}_{ph} - wt)]. \qquad (2)$$

Here $L$ is the normalization length of the photon wavefunction, $\boldsymbol{r}_{ph}$ is the coordinate vector of the photon with frequency $w = ck$, wavevector $\boldsymbol{k}$ and polarization $s$, $\boldsymbol{e}_{\boldsymbol{k},s}$ are polarization vectors normal to $\boldsymbol{k}$ and $C_{\boldsymbol{k},s}$ is the photon wavefunction in the momentum representation. Implicit in the definition of the photon wavefunction is that the photon, seen as a concentration of energy that can be localized up to the limit imposed by the uncertainty principle between the electric and magnetic field operators (or between the annihilation and creation operators), propagates with the light velocity. The massless photon cannot interact with the environment (such an interaction, expressed in classical mechanics through a force and in quantum mechanics through a change in energy, would require a finite mass), but can only be emitted or absorbed by the environment.

What happens if the electrically charged material system is in its ground state? A quantum particle, whether in its excited or ground state, cannot be at rest (or in uniform motion) due to the uncertainty principle between the coordinate and momentum operators, and therefore, if electrically charged, it must emit radiation. Classical mechanics tells us that an accelerated charge emits radiation until it arrives in the state of rest, but in quantum mechanics, since the rest state is forbidden, the radiated energy of the particle in the ground state is recuperated from the quantum vacuum. This mainstream interpretation of the stability of quantum systems assumes that the quantum vacuum exists independent of the material system. Due to the controversies related to this interpretation, we look for an alternative explanation of the stability of quantum systems.

Such an explanation can be found supposing that there is no fundamental difference between the material particle in its excited and ground state; the equation of motion is the same in both cases. Then, it follows that the quantum harmonic oscillator in the ground state should emit a photon in decaying to a lower energy state, just as it does in the excited state. As a result, the energy of the harmonic oscillator becomes negative, equal to $-\hbar w_0/2$. Figure 1 shows schematically the transition of the electron from the ground state $|\Psi_+, \boldsymbol{p}\rangle = |\Psi_+\rangle|\boldsymbol{p}\rangle$, where $|\Psi_+\rangle$ is the positive ground state of the harmonic oscillator with center-of-mass energy $E_+ = \hbar w_0/2$ and $\boldsymbol{p}$ is the momentum of the particle, into the negative-energy state $|\Psi_-, \boldsymbol{q}\rangle = |\Psi_-\rangle|\boldsymbol{q}\rangle$, where $|\Psi_-\rangle$ is the state of the harmonic oscillator with center-of-mass energy $E_- = -\hbar w_0/2$ and $\boldsymbol{q}$ the corresponding momentum, accompanied by the emission of a photon with energy $\hbar w$ and wavevector $\boldsymbol{k}$.

Formally, a negative-energy quantum state of a harmonic oscillator with mass $m$ can be interpreted as the state of a harmonic oscillator with mass $-m$. It is clear that if

$$\hat{\boldsymbol{H}}_+|\Psi_+\rangle = (\hat{\boldsymbol{p}}^2/2m + m\boldsymbol{w}_0^2\hat{\boldsymbol{r}}^2/2)|\Psi_+\rangle = E_+|\Psi_+\rangle, \tag{3}$$

then

$$\hat{\boldsymbol{H}}_-|\Psi_+\rangle = (-\hat{\boldsymbol{p}}^2/2m - m\boldsymbol{w}_0^2\hat{\boldsymbol{r}}^2/2)|\Psi_+\rangle = -E_+|\Psi_+\rangle = E_-|\Psi_+\rangle, \tag{4}$$

so that $|\Psi_+\rangle = |\Psi_-\rangle$. The quantum state of a negative-mass harmonic oscillator is the same as that of a positive-mass harmonic oscillator with the same parameters! In quantum optics negative energy states of photons are simply considered as states with opposite helicity [2]. Nevertheless, negative-energy states of material particles are not commonly encountered in quantum mechanics and for a very good reason: they are unstable. The cause of their instability can be qualitatively understood through the following argument: just as an accelerated charged particle with positive mass radiates energy, an accelerated charged particle with negative mass absorbs energy (the equations of motions are the same for the two situations). For an isolated particle the only energy that can be absorbed when it jumps into the state with negative energy is the photon emitted during that transition. Therefore, the particle in the state $|\Psi_-\rangle$ absorbs the photon, jumps into the state $|\Psi_+\rangle$, which then decays into the state $|\Psi_-\rangle$ and a photon, and so on.

The unstable quantum particle exists in the state with negative energy for a time interval $t$ given by the uncertainty condition: $(E_+ - E_-)t \approx 1$, and the accompanying photon is considered to be real, not virtual. The denomination of quantum vacuum is, however, appropriate even in this case since the photon does not become separated from the material particle; they evolve in an entangled state for a time $t$. But, since it exists for the time $t$, the electromagnetic field state extends around the material particle over a distance $ct \approx \lambda/2\pi$, with $\lambda$ the wavelength of the emitted radiation, if we assume the low-energy interaction regime $\hbar\omega/mc^2 \ll 1$, for which $\omega \cong \omega_0$. In this regime, the validity of which is supposed throughout this paper, the kinetic energies of the material particle in the initial and final states are negligible in comparison with the photon energy. The distance of photon propagation is larger for lower-energy transitions; effects related to quantum vacuum are primarily low frequency, non-relativistic effects. Although the universe is not bathed in ZPE we still sense it around us because it is present around any material particle.

Throughout the photon emission and re-absorption processes the energy of the compound system is equal to that of the material particle in its positive-energy ground state, the difference from standard quantum theory being that the quantum vacuum is not considered as existing independent of the material particle, but as generated by it, and so as localized around the material particle. The apparent energy of the ZPE can be taken as half of that of the photon emitted during transitions to the negative-energy states, i.e. equal to the usual expression for the ZPE.

It is important to emphasize that the negative mass is dynamic in character; no negative gravitational mass is involved, and no negative inertial mass is encountered because in the negative-energy state the material particle is entangled with the photon and hence no force (interaction, in general) applies uniquely on the particle. Since the photon is afterwards re-absorbed, it is not spatially separated enough from the negative-energy particle in order to assure localized interactions of the negative-energy particle alone. The entangled state is a consequence of energy- and momentum-conservation laws at transition, as for the case of the excited-ground state transition in [3].

It is interesting to point out that the interpretation of the ZPE and the relating effects (Casimir force, Lamb shift, and so on) as being entirely due to either the vacuum field or the material system or both depending on the ordering of operators gains a new physical insight in the present interpretation of quantum vacuum: the ZPE cannot indeed be associated to either the localized vacuum nor to the material system. Its existence reflects the entangled state between photons and the material particle with negative energy. However, this entangled state cannot be separated into a photon and a negative energy particle and so for all practical purposes the standard quantum theory of a particle in the ground state in interaction with the (now localized and dependent on the particle) quantum vacuum remains valid. As in standard quantum theory, the expectation values of the electric and magnetic fields in the vacuum states vanish since the direction of photon emission is random. In its turn, the random particle recoil, over distances comparable to the Compton wavelength of the material particle, can be viewed as a form of Zitterbewegung (for discussions on this subject see [8,9] and the references therein). Since the photon remains entangled with the particle during its existence, it is expected that in this mechanism of Zitterbewegung the random particle recoils are pair wise correlated: the particle recoils at emission and absorption of the same photon should be identical.

The Zitterbewegung in our model is a consequence of photon emission and re-absorption during transitions from positive- to negative-energy states. These transitions take place for a non-relativistic electron. Note that in the relativistic Dirac electron the Zitterbewegung is similarly considered as due to virtual, non-energy-conserving transitions during which positive and negative energy electrons in the Dirac sea exchange roles (see [1] for an insightful treatment of the subject). The mechanism of Zitterbewegung proposed in the present paper involves real, photon-mediated and energy-conserving transitions between positive- and negative-energy states, and explains both the origin of the ZPE and the cause of non-observing negative-energy states in quantum mechanics: they cannot be perceived as such because they cannot be disentangled from the accompanying photon.

For a better understanding of the behavior of a material particle in the negative-energy state it is illuminating to compare in more detail the photon-assisted transitions of a material particle between positive-energy states and between states with opposite energies. The first major difference is that in the second case the particle recoil has the same direction as the emitted photon. For example, in a coordinate system in which the initial positive-energy state particle is at rest conservation of momentum requires that $p = 0 = q + \hbar k$, with $q$ the momentum of the material particle in the negative-energy state. Then, since $q = -mv$, it follows that the negative-energy particle recoils in the same direction as the emitted photon!

This certainly contradicts common sense (supported also by experiments), which tells us that the material particle recoils in a direction opposite to that of the emitted photon, but we are dealing with a negative-mass particle, which is not in itself a common-sense concept since it cannot be detected separately.

A more thorough comparison of the photon–particle entangled states in the situations of photon emission from the excited and ground state can be performed following the mathematical treatment of photon–atom entanglement in [3]. Reference [3] deals with spontaneous emission of a photon from a finite-size wave-packet representing an excited atom that undergoes a transition to its ground state, and finds the analytical expression for the atom–photon entangled state in both momentum and coordinate representations when recoil is taken into account. We consider here a similar problem to that in [3], namely the spontaneous emission of a photon during particle transition from the positive-energy ground state to a negative-energy state, the initial state in our case having the same characteristics as the excited state in [3]. The only differences from [3] are that in our case the particle in the negative-energy state has a negative mass and that the time coordinate is limited to $t$; these differences lead, as we show in the following, to a particle behavior that is qualitatively different than in [3]. For ease of comparing the final results, we follow closely the notations in [3] and, in particular, put $\hbar = 1$ in the subsequent calculations. Let us denote by

$$|\Psi\rangle = \sum_{q,k} C_{q,k}(t) \exp[-i(E_- + w - q^2/2m)] |\Psi_-, q\rangle |1_k\rangle \qquad (5)$$

the entangled state between the emitted photon and the recoiled quantum particle in the negative-energy state, with kinetic energy $-q^2/2m$ (the kinetic energy of the particle in the positive-energy state is $(q+k)^2/2m$), and with $\Psi(r_{at}, r_{ph}, t)$ the corresponding wavefunction in the coordinate representation:

$$\Psi(\mathbf{r}_{at}, \mathbf{r}_{ph}, t) = L^{-3} \sum_{\mathbf{q},\mathbf{k}} C_{\mathbf{q},\mathbf{k}}(t) \mathbf{e_k} \exp[i(\mathbf{q} \cdot \mathbf{r}_{at} + \mathbf{k} \cdot \mathbf{r}_{ph})] \exp[-i(E_- + \mathbf{w} - q^2/2m)t]. \qquad (6)$$

Here $\mathbf{r}_{at}$ is the position vector of the material particle, and we assume, as in [3], that only photons with a given polarization vector $\mathbf{e_k} = [k^2 \hat{\mathbf{z}} - \mathbf{k}(\mathbf{k} \cdot \hat{\mathbf{z}})]/[k\sqrt{k^2 - (\mathbf{k} \cdot \hat{\mathbf{z}})^2}]$ in the $(\mathbf{k}, \hat{\mathbf{z}})$ plane are emitted, where $z$ is the intra-atomic electron coordinate and $\hat{\mathbf{z}}$ is the unit vector along $z$. If, moreover, in the ground state $|\Psi_+\rangle$ the particle center-of-mass wavefunction in the coordinate representation, $\Psi_+(\mathbf{r}_{at}, t)$, has an initial Gaussian form:

$$\Psi_+(\mathbf{r}_{at}, t=0) = (\sqrt{\mathbf{p}} a_0)^{-3/2} \exp(-r_{at}^2/2a_0^2) \qquad (7)$$

the expansion coefficients $C_{\mathbf{q},\mathbf{k}}$ in the long-time limit $t \gg 1/\mathbf{g}$, with $\mathbf{g}$ the decay rate, are given by [3]

$$C_{\mathbf{q},\mathbf{k}}(t) = -\frac{iez_{+-}\mathbf{w}_0(2\mathbf{p})^2}{\sqrt{\mathbf{w}}} \left(\frac{a_0}{L^2\sqrt{\mathbf{p}}}\right)^{3/2} \frac{\sin \mathbf{q}}{-q^2/2m - (\mathbf{q}+\mathbf{k})^2/2m + \mathbf{w} - \mathbf{w}_0 + i\mathbf{g}/2}. \qquad (8)$$

In (8) $z_{+-}$ is the matrix element of the positive–negative-energy state transition, and $\mathbf{q}$ the angle between the $z$ axis and $\mathbf{k}$. We mention that the long-time limit in our case has a slightly different meaning than in [3]: in [3] it implies that all excited atomic population returns to the ground state, whereas here it simply indicates that the single quantum particle has jumped to the negative-energy state. Of course, we assume that $t \gg 1/\mathbf{g}$.

A first integration of (6) over the solid angle element in the $\mathbf{k}$ direction made under the assumptions that $\mathbf{w}_0/mc^2 \ll 1$ (and hence $|\mathbf{k}| \cong \mathbf{w}_0/c$) and that the relative-motion position vector $\tilde{\mathbf{n}} = \mathbf{r}_{ph} - \mathbf{r}_{at}$ is parallel to $\mathbf{k}$, leads, similar to equation (19) in [3], at the expression

$$\Psi(\mathbf{r}_{at},\mathbf{r}_{ph},t) = -\left(\frac{ez_{+-}w_0 a_0^{3/2} \exp(-iE_- t + iw_0^2 t/2mc^2)}{(2p)^3 p^{3/4} c^{1/2}}\right)\frac{\mathbf{e}'\sin q'}{r}$$

$$\times \int d\mathbf{q} \exp\left[-\frac{q^2(a_0^2 - it/m)}{2} + i\left(\mathbf{q}\cdot\mathbf{r}_{at} - \frac{tq'w_0}{mc}\right)\right]\int dk \frac{k^{1/2}\exp[ik(r-ct)]}{-q^2/m - w_0^2/2mc^2 + q'w_0/mc + w - w_0 + ig/2}$$

(9)

Here $q' = \mathbf{q}\cdot\mathbf{k}/k$, $\mathbf{e}'$ is a unit vector normal to $\mathbf{r}$ and lying in the $(\hat{z}, \tilde{\mathbf{n}})$ plane, and $q'$ is the angle between $\hat{z}$ and $\mathbf{r}$. Note that in [3] the hypothesis $\mathbf{k} \parallel \tilde{\mathbf{n}}$ followed from the far-zone approximation $kr \gg 1$, whereas in our case we assume that $\tilde{\mathbf{n}}(t) = \tilde{\mathbf{n}}(0) + (c\mathbf{k}/k + \mathbf{q}/m)t$ $\cong \tilde{\mathbf{n}}(0) + ct\mathbf{k}/k \cong ct\mathbf{k}/k$. This approximation holds for most of the time in which the emitted photon exists (the inequality $w_0/mc^2 \ll 1$ was again employed when dropping the $\mathbf{q}/m$ term), as long as the initial $\mathbf{r}$ value is within the Compton wavelength of the material particle. After performing the integral over $k$, the entangled wavefunction, analogous to [3], is given up to phase factors by the following expression:

$$\Psi(\mathbf{r}_{at},\mathbf{r}_{ph},t) \approx \left(\frac{ez_{+-}w_0^{3/2} a_0^{3/2}}{(2p)^2 p^{3/4} c^2}\right)\frac{\mathbf{e}'\sin q'}{r}\Theta(ct-r)\exp\left(\frac{g}{2c}(r-ct)\right)$$
$$\times \int d\mathbf{q} \exp\left[-\frac{q^2}{2}\left(a_0^2 + i\frac{(t-2r/c)}{m}\right) + i\mathbf{q}\cdot\mathbf{R}\right]$$

(10)

where $\mathbf{R} = \mathbf{r}_{at} - \tilde{\mathbf{n}}w_0/mc^2$ is the center-of-mass position vector and $\Theta$ is the step function. Note that (10) differs from the corresponding equation (21) in [3] through an additional $\mathbf{r}$-dependent contribution to the imaginary term that multiplies $q^2$ in the argument of the exponential function in the integral over $\mathbf{q}$, which derives from the different denominator in (8) as compared to the corresponding equation (15) in [3]; this difference in its turn originates from the negative mass in the lower-energy state as compared to the positive mass in [3].

Finally, after performing the remaining integral, the entangled wavefunction is

$$\Psi(\mathbf{r}_{at},\mathbf{r}_{ph},t) \approx \frac{ez_{+-}w_0^{3/2}}{(2\pi)^{1/2}p^{3/4}c^2}\frac{e'\sin q'}{r}\Theta(ct-r)\exp\left(\frac{g}{2c}(r-ct)\right)$$
$$\times \frac{1}{[a_0+i(t-2r/c)/ma_0]^{3/2}}\exp\left(-\frac{R^2}{2[a_0^2+i(t-2r/c)/m]}\right), \qquad (11)$$

its squared modulus $|\Psi(\mathbf{r}_{at},\mathbf{r}_{ph},t)|^2 = |\Psi_{rel}(\tilde{\mathbf{n}},t)|^2 |\Psi_{cm}(\mathbf{R},\tilde{\mathbf{n}},t)|^2$ being separable into a relative-motion wavefunction

$$|\Psi_{rel}(\tilde{\mathbf{n}},t)|^2 = \frac{3g\sin^2 q'}{8\pi c r^2}\Theta(ct-r)\exp\left(\frac{g}{c}(r-ct)\right) \qquad (12)$$

that has the same entanglement-free photon wavefunction form as in [3] and a term

$$|\Psi_{cm}(\mathbf{R},\tilde{\mathbf{n}},t)|^2 = [\sqrt{\pi}\,a(\mathbf{r},t)]^{-3}\exp[-R^2/a(\mathbf{r},t)^2] \qquad (13)$$

that resembles a center-of-mass particle wavefunction with a time- and $\mathbf{r}$-dependent width $a(\mathbf{r},t) = [a_0^2 + (t-2r/c)^2/m^2 a_0^2]^{1/2}$.

The calculation of the entangled negative-energy particle–photon state performed with the formalism in [3] shows that the entangled state in our case differs from that in [3] in three important respects:

(i) $|\Psi_{cm}(\mathbf{R},\tilde{\mathbf{n}},t)|^2$ depends also on $\mathbf{r}$, not only on $\mathbf{R}$, which means that it has not the form of an entanglement-free center-of-mass particle wavefunction.

(ii) the width $a(\mathbf{r},t)$ of the Gaussian center-of-mass wavefunction is not equal to $a_0$ even when $t = 0$. This discontinuity is related to the creation of the particle–photon pair and can be considered as a mathematical expression of the related Zitterbewegung. The sudden increase of the width of the entangled wavefunction $a(\mathbf{r},t)$ at $t = 0$ is within the Compton wavelength.

(iii) $a(\mathbf{r},t)$ does not increase uniformly with time. Indeed, after the initial jump in $a(\mathbf{r},t)$ associated to the Zitterbewegung, the width of the center-of-mass part of the entangled wavefunction initially decreases in time, attaining the $a_0$ value after $t = 2\mathbf{r}/c$ and then increases again up to a value approximately given by the expression in [3] for $t = \mathbf{t}$. The width of the entangled wavefunction, given by $a(\mathbf{r}, t \cong \mathbf{t}) = [a_0^2 + (\mathbf{r}/c)^2/m^2 a_0^2]^{1/2}$ for $t \cong \mathbf{t}$, decreases to the initial value $a_0$ when the photon is re-absorbed and the particle returns to the positive-energy ground state since in this case we can formally set $\mathbf{r} = 0$. The cycle can start again.

The conclusion is that entangled photon–negative-energy states differ qualitatively from entangled photon–positive-energy states. The creation of such entangled states by photon emission from the ground state of a material particle is associated with a sudden change in the width of the center-of-mass part of the wavefunction, which can be interpreted as a signature of Zitterbewegung. The entangled photon–positive-energy state exists only for a limited time interval, after which the unstable negative-energy particle absorbs the photon and performs an energy- and momentum-conserving transition into the ground state. The generation of a photon during a limited time interval, with a random wavevector, accounts for a quantum vacuum state localized around the material particle, with a ZPE given by the expression in standard quantum theory.

How does the present interpretation of the origin of ZPE change the predictions of standard quantum theory? As regards the quantum vacuum effects that take place in or around material systems (for example, Lamb shift, vacuum polarization near charged particles [1]), the predictions do not change since these effects occur themselves inside or near material systems. The existence of the (static and dynamic) Casimir effect and the associated phenomena (dependence of field commutators on boundaries, the predicted but not yet observed emission of photons from vacuum in the neighborhood of moving boundaries or time-varying dielectric constants, as summarized in [6], sonoluminescence [10], etc.) is also not endangered by this interpretation of ZPE, its presence for very small distances between nano-sized objects being in fact outlined by the present explanation. A closer look to a standard derivation of the Casimir force between parallel conducting plates, for example (see [1]), reveals that the expression of the Casimir force remains the same. Actually, in any derivation of this force an arbitrary frequency cut-off is introduced (the Casimir force does not depend on this cut-off), and the quantization of the electromagnetic field wavevectors in the space between the plates is employed. Both these essential features are still valid in the present interpretation of the quantum vacuum, with the added insight that the frequency cut-off corresponds to the highest oscillation frequency of the material system. Since this highest oscillation frequency is finite for any particle, the energy density of the vacuum state cannot go to infinity even in the neighborhood of material systems. And, of course, there should be no cosmological paradox since the ZPE does not exist far from matter; it does not overflow the universe, although cosmological effects related to the quantum fields exist. Actually, the spectacular cosmological effect of vacuum lensing [11] due to light propagation in the neighborhood of magnetized neutron stars, which influence the quantum vacuum through their magnetic field, can be easily grasped in the present interpretation of quantum vacuum.

DISCUSSIONS AND CONCLUSIONS

In the previous section we have provided a model for the quantum vacuum that apparently removes the difficulties associated to the ZPE in the standard quantum field theory. But are there more direct arguments for or against a localized ZPE?

Besides the Casimir effect, statistical mixtures of quantum vacuum and single-photon Fock states have been generated in [12], in an experiment that does not seemingly involve material systems around which the ZPE is located. This experiment and others of the same type appear to indicate that ZPE can be separated from matter. However, such a conclusion is not straightforward, since experiments of this type do not measure directly the ZPE (the ZPE has no photons that can be counted), but measure noises in different photodetectors. It is not possible to discern between the case when the ZPE exists and propagates independently of the photodetectors and the case when it exists around the photodetectors because its is generated by these material systems; in both cases the photodetectors would give the same answer and hence no conclusion of the independent existence of ZPE can be drawn from noise measurements. Similarly, squeezing of quantum vacuum cannot be considered an argument for its existence independent of material system, since such squeezings are usually done via interactions with material systems (for example, via polarization self-rotation in rubidium vapors in [13]).

Another quantum vacuum effect that is not manifestly related to material systems is vacuum birefringence or vacuum polarization in the presence of strong electromagnetic fields [14]. This effect has not yet been demonstrated experimentally [15], and our model predicts that it can only occur in the neighborhood of material systems.

Ideas similar to those presented here have appeared in different contexts. For example, in [16] it was argued that the contribution of vacuum fluctuations to the Casimir effect and spontaneous emission of radiation can be understood in terms of virtual photon creation at the position of charged particles. The creation of such a localized photon implies that the associated wavevectors and polarizations can have arbitrary values, simulating the fluctuations of an infinitely extended vacuum field; the hypothesis of an extended ZPE with a divergent energy density is not necessary. The difference in the present model is that we consider the photon creation as a real process, fact that also allows an explanation of the Zitterbewegung and the absence of negative energy states in quantum mechanics. On the other hand, we don't believe that a photon can be created at a precise location since localization of a quantum particle is forbidden by the uncertainty principle.

Another approach towards demonstrating the finiteness of ZPE density has been taken in [17]. A finite ZPE density can in principle be found in a loopwise summation procedure accompanied by a renormalization-group analysis, if an eigenvalue condition is imposed on the renormalized fine-structure constant. Unfortunately, due to computational difficulties no explicit solution to the imposed condition has been found.

Indirect support to our model comes for the work in [3,18], in which it is shown that spontaneous emission of photons from an excited atom in free space, associated with recoil, entangles the momenta of the recoiling atom and the photon. Such an entanglement is also the basis of our model, with the difference that the recoiled particle has a negative mass and is therefore unstable, so that the entangled state has a finite lifetime.

In conclusion, although there is no experimental observation to support our theory, there is none to infirm it either. From a conceptual point of view, we believe that the model of quantum vacuum presented in this paper has a number of advantages compared to other approaches of the same problem. Among these, the model describes in a unified manner the stability of quantum material systems, the reality of the Casimir and other quantum vacuum effects predicted to occur in the neighborhood of materials systems, the Zitterbewegung, and the existence of a ZPE that is not extended throughout the space and hence does not contradict the observed value of the cosmological constant. Last, but not least, the model of photons and the ZPE in this approach does not make use of the harmonic oscillator analogy, which contradicts the relativity theory in that it applies a mathematical formalism designed for material systems to massless particles (photons). The model of quantum vacuum presented in this paper predicts that the Zitterbewegung is pair wise correlated and that vacuum birefringence cannot be observed far from material systems.

FIGURE CAPTIONS

Fig.1 Schematic representations of the sequence of transitions from the positive-energy ground state of the material particle to a negative-energy state and a photon, which assures the stability of quantum systems

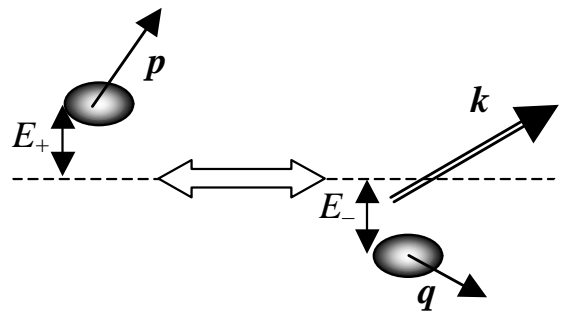

Figure 1

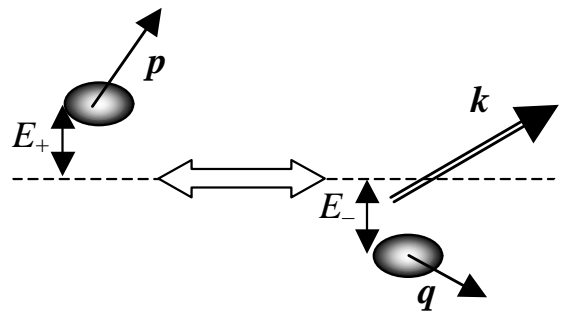

Figure 1